\documentclass[aps,preprint,showpacs,flotfix]{revtex4-2}
\usepackage{latexsym}
\usepackage{graphics}
\usepackage{graphicx,amssymb,epsfig,epsf,amsmath}
\usepackage{physics}
\usepackage[colorlinks]{hyperref}
\usepackage{multirow}

\def\be{\begin{equation}}

	\def\ee{\end{equation}}
\def\barr{\begin{array}}
	\def\earr{\end{array}}

\def\nn8{\\}
\def\l{\left}
\def\r{\right}
\def\dis{\displaystyle}
\def\ed{\end{document}}

\def\cac{{\cal C}}

\def\can{{\cal N}}

\begin{document}

\title{Average-fluctuation separation in energy levels in many-particle quantum systems with $k$-body interactions\\ using $q$-Hermite polynomials}
\author{N. D. Chavda\footnote{ndchavda-apphy@msubaroda.ac.in}}
\affiliation{Department of Applied Physics, Faculty of Technology and Engineering,\\ The Maharaja Sayajirao University of Baroda, Vadodara-390001, India}




		
\begin{abstract}
	
Separation between average and fluctuation parts in the state density in many-particle quantum systems with $k$-body interactions, modeled by the
$k$-body embedded Gaussian orthogonal random matrices (EGOE($k$)), is demonstrated using the method of normal mode decomposition of the spectra and also verified through power spectrum analysis, for both fermions and bosons. The smoothed state density is represented by the $q$-normal distribution ($f_{qN}$) (with corrections) which is the weight function for $q$-Hermite polynomials. As the rank of interaction $k$ increases, the fluctuations set in with smaller order of corrections in the smooth state density. They are found to be of GOE type, for all $k$ values, for both fermion and boson systems.

\end{abstract}





\maketitle

\section{Introduction}

The understanding of separation of information, in energy levels as well as in other observables, into averages (smooth part) and fluctuations, in finite interacting quantum systems like nuclei, atoms and molecules is very important as it forms a physical basis for the statistical spectroscopy \cite{Brody81}. With this, the average properties can be studied following the spectral distribution methods developed by French and co-workers \cite{FK1982,FRSHK06,Karw94,FGGP99,AK05,kota-book}, while the fluctuations can be analyzed following the classical Gaussian orthogonal ensemble (GOE) or Gaussian unitary ensemble (GUE)
or Gaussian symplectic ensemble (GSE) of random matrix theory (RMT)  depending on the symmetries possessed by the quantum systems \cite{Wig1955,Bohrev,Stoc,Casati1980,Haake2010}. To confirm GOE fluctuations, one generally uses measures like the nearest neighbor spacing distribution (NNSD), giving the degree of level repulsion, and the Dyson-Mehta ($\Delta_3$) statistic \cite{Dyson} or the power spectrum, giving the long-range spectral rigidity. In order to construct these measures for a given set of energy levels, one has to renormalize the eigen-energies to remove the secular variation in the state density. This method is called unfolding of the spectra \cite{Brody81,Haake2010} in which fluctuations are separated out by removing the smooth or average part from the spectra and it can be done in two different ways: If the smooth form of the eigenvalue density is already known then it is possible to expand state density in terms of its asymptotic (or smoothed) form using the orthonormal polynomials defined by the asymptotic density. On the other hand, when the form of eigenvalue density is not known then one uses polynomial unfolding.

In last couple of decades, there is tremendous growth on the study of statistical properties of isolated finite many-particle quantum systems such as atomic nuclei, atoms, quantum dots and small metallic grains, ultracold atoms, interacting spin models, and quantum black holes with the Sachdev-Ye-Kitaev (SYK) model and so on \cite{kota-book,pol2011,LYAM2016,BISZ2016,KC2018,KC2018a,Cotler17,Benet2001,PW2007,Gracia16-17,Jia2020}. The embedded ensembles of $k$-body interaction, EGOE($k$), operating in many particle spaces \cite{Brody81,kota-book}, provide the generic models for finite interacting many-particle systems as inter-particle interactions are known to be predominantly few-body in character. With $k=2$, called two-body random ensembles, EGOE(2), were first introduced in the context of the nuclear shell model \cite{FW1970,BF71}. These models for the particles in a mean-field and interacting via two-body interactions, and their different extended versions are analyzed, in detail for both, fermionic systems \cite{Brody81,FGT1996,FI1997,MF1975,kota2001} as well as bosonic systems \cite{PDPK2000,ABRW20012002,CPK20032004,CKP2012}. One of the most significant features of EGOE(2) is the well defined separation of information, in energy levels (and also  in observables), into averages and fluctuations. The nature of this separation is quite well understood analytically using so called the binary correlation approximation for fermionic EGOE(2) in the dilute limit ($m \rightarrow \infty$, $N \rightarrow \infty$, $m/N \rightarrow 0$) \cite{Brody81,MF1975} and also for bosonic BEGOE(2) (B for boson) in the dense limit ($m \rightarrow \infty$, $N \rightarrow \infty$, $m/N \rightarrow \infty$) \cite{PDPK2000,Lec2008}, where $m$ represents fermions/bosons in $N$ single particle (sp) states. Numerical examples are also given in \cite{PDPK2000,Lec2008,Lab90}.

Going beyond two-body interactions, the importance of embedded ensembles of $k$-body interaction, EGOE($k$), is renewed further in the series of recent investigations employing $q$-Hermite polynomials \cite{Vyas2019,KM2020,rao21,KM2021}. The motivation is mainly followed from the work by Verbaarschot and collaborators, who have employed the weight function for $q$-Hermite polynomials, so called $q$-normal distribution ($f_{qN}(x|q)$), to study spectral densities in quantum black holes with Majorana fermions \cite{Gracia16-17,Jia2020}. It is worth recalling  that the interactions with higher body rank $k>2$ are very important in nuclear physics \cite{LaDyDra2012}, strongly interacting quantum systems \cite{Bla-Mc75,HaNoS2013}, quantum wormholes \cite{GarNoRoVer2019} and black holes \cite{Cotler17,GarJVer2018} with Majorana fermions and in disordered finite quantum networks \cite{Ortega2015,Ortega}.

Embedded ensembles for $k$-body interactions are studied for both fermion and boson systems. It is demonstrated that the smooth form of the eigenvalue density, generated by EGOE($k$) for both fermions and bosons (and also corresponding EGUE($k$) versions), is represented by $f_{qN}(x|q)$ and it correctly describes the transition in spectral density from Gaussian (for parameter value $q=1$) to semi-circle (for $q=0$) as the rank of interaction $k$ changes
from $1$ to $m$ \cite{Vyas2019}. Importantly in recent analysis using one- plus random $k$-body interactions for interacting dense boson systems \cite{rao21}, it is shown numerically that $f_{qN}(x|q)$ can also represent the smooth form of the state density as the strength of $k$-body interaction, $\lambda$, varies. Further, the formula for onset of thermalization ($\lambda_t$) in dense boson systems and the smooth forms for the number of principal components, valid for all $k$ and $\lambda > \lambda_t$, are derived and well verified with ensemble averaged numerical results \cite{rao21}. It is important to note that the derivation of these smooth forms essentially require the level fluctuations as well as the strength fluctuations to be of GOE type \cite{KS2001}. In the past, French and co-workers \cite{Brody81,MF1975}, using EGOE(2) and nuclear shell model examples, conjectured that the level and strength fluctuations follow GOE for EGOE($k$) and thereafter many attempts were made to establish the nature of fluctuations generated by these ensembles \cite{Benet2001,verb1984,Papen2011}. With replica trick developed in statistical mechanics for the study of spin glasses and Anderson localization, the two-point function for EGOE($k$) was derived in \cite{verb1984}. Further, using properties of the expansion and the supersymmetry technique in \cite{Benet2001}, it was shown that for $k>m/2$, smooth part of the spectrum has shape of a semicircle with the spectral fluctuations are of GOE type. With an analogy to the metal-insulator transition in \cite{Papen2011}, the nature of spectral fluctuations generated by finite Fermi systems modeled by a few-body interaction, in presence of a mean field, were shown to be of GOE type. More recently, using the normal mode decomposition of the spectral density of SYK model, the fluctuations are analyzed via two-point function using $q$-Hermite orthogonal polynomials and demonstrated that only a small number of polynomials are sufficient for a separation of scales between long-wavelength fluctuations of the spectral density and the short-wavelength fluctuations of the universal RMT spectral correlations \cite{Jia2020}. Employing EGOE($k$) for fermions and bosons, the main purpose of the present study is to establish that finite interacting many-particle quantum systems with $k$-body interactions exhibit average-fluctuations separation in energy levels with smooth density given by $f_{qN}(x|q)$ and the fluctuations are of GOE type.

The paper is organized as follows:  The embedded ensembles for interacting quantum systems are briefly defined in Section~\ref{sec:2}. Some basic properties of $q$-Hermite polynomials and the univariate $q$-normal distributions $f_{qN}(x|q)$ are defined in Section~\ref{sec:3}. Results
for level motion in dilute fermion and dense boson systems with body rank $k$, obtained via the binary correlation method are discussed briefly in Sections~\ref{sec:4}. Numerical results for the normal-mode decomposition of the spectra, for various values of $k$, along with the results of periodogram analysis of average-fluctuation separation, are presented in Section~\ref{sec:5}. Also Section~\ref{sec:5} includes results for NNSD and $\Delta_3$ statistics for these examples. Finally, Section~\ref{sec:6} gives concluding remarks.
	
\section{Embedded Ensembles}
\label{sec:2}
Consider $m$ spin-less fermions (bosons) occupying $N$ sp states and interacting via $k$-body interaction ($2 \leq k \leq m$). Let these $N$ sp states be denoted by $|v_{i} \rangle$, where $i=1,2,3,\ldots,N$ with $\sum v_i=k$. The number
of states in $m$-particle system corresponding to the dimension of Hamiltonian matrix $d(N,m)=\binom{N}{m}$ for fermions (and $d(N,m) = \binom{N+m-1}{m}$ for bosons). A basis state in $m$-particle space is denoted by $|\prod_{i=1}^{N} m_i \rangle$ with $\sum m_i=m$. Using a GOE representation for the $k$-particle Hamiltonian matrix and then propagating the information into $m$-particle spaces using the concepts of direct product space and Lie algebra, one obtains EGOE($k$) for fermions (and BEGOE($k$) for bosons) \cite{Ortega2015,Vyas2019}. For such a system, the $k$-body Hamiltonian matrix is defined as follows:
\begin{equation}
V(k)= \sum_{\alpha,\gamma} V_{k:\alpha,\gamma} \ A_{k,\alpha}^{\dag} A_{k,\gamma}\;.
\label{H}
\end{equation}
Here, $\alpha$ and $\gamma$ denote $k$-particle configuration states in occupation number basis and $V_{k:\alpha,\gamma}$'s  are the Gaussian random variables
with,
\begin{equation}
\overline{V_{k:\alpha,\gamma}} = 0\;,\;\;\;\;
\overline{V_{k:\alpha,\gamma} V_{k:\alpha',\gamma'}} = \nu^{2} (\delta_{\alpha,\alpha'} \delta_{\gamma,\gamma'} +\delta_{\alpha,\gamma'} \delta_{\alpha',\gamma})\;,
\label{H_1}
\end{equation}
where the bar indicates ensemble average and $\nu^2=1$ is used in the present analysis. For fermions,
\[ A_{k,\alpha}^{\dag} = \prod_{i=1}^{k} f_{v_{i}}^{\dag} ; A_{k,\alpha} = (A_{k,\alpha}^{\dag})^{\dag} \;( v_{1} <v_{2}<\ldots < v_{k} ),\]
where $f^{\dag}_{v_{i}}$ and $f_{v_{i}}$ are the creation and annihilation operators respectively. For bosons, 
\[ A_{k,\alpha}^{\dag} = {\cal N} \prod_{i=1}^{k} \ b_{v_{i}}^{\dag};  A_{k,\alpha} = (A_{k,\alpha}^{\dag})^{\dag} \;( v_{1} \leq v_{2} \leq \ldots \leq v_{k} ),\]
where $b^{\dag}_{v_{i}}$ and $b_{v_{i}}$ are the creation and annihilation operators respectively. The normalization factor ${\cal N}$, normalizes $k$-particle bosonic states and $\alpha$ simplifies the notation of indices. One should also note that dimension of $k$-particle Hamiltonian is $d(N,k)$ with $k$-body independent matrix elements (KBME) being $\frac{d(N,k)(d(N,k)+1)}{2}$. $V_{k:\alpha,\gamma}$ in Eq.\eqref{H}, are anti-symmetrized KBME for fermions (and symmetrized for bosons). For $k=m$, EGOE($k$) and BEGOE($k$) are identical to GOE with dimensionality $d(N,m)$.

\section{$q$-Hermite polynomials}
\label{sec:3}

Let us start with $q$-Hermite polynomials, introduced first
in Mathematics to prove the Rogers–Ramanujan identities \cite{Ismail}, defined by the following recurrence relation \cite{Ismail,Szab}:

\be
	\begin{array}{rcl}
x H_n(x|q) &=& H_{n+1}(x|q) + [n]_q H_{n-1}(x|q);\\\\
&&H_0(x|q)=1, H_{-1}(x|q)=0.
	\end{array}
\ee
Here, with $q$ as parameter, $q$ numbers, denoted by $[n]_q$, are defined as,
\[ [n]_q = 1+q+q^2+\ldots+q^{n-1} = \frac{1-q^n}{1-q}. \] Then as $q\rightarrow1$, $[n]_q = n$, and $q$ factorial of $n$, $[n]_q ! = \prod_{j=1}^{n} [j]_q$ with $[0]_q! = 1$. The formulas for $H_n(x|q)$ upto  $n=6$ are given below:
\begin{equation}
	\begin{array}{rcl}
		H_1(x|q) &=& x \;,\\
		H_2(x|q) &=& x^2-1 \;,\\
		H_3(x|q) &=& x^3 - (2 + q) x \;,\\
		H_4(x|q) &=& x^4 - (3 + 2q + q^2) x^2 + (1 + q + q^2)\;,\\
		H_5(x|q) &=& x^5 - (4 + 3q + 2q^2 + q^3) x^3 \\
&& + (3 + 4q + 4q^2 + 3q^3 + q^4) x\;,\\
		H_6(x|q) &=& x^6 - (5 + 4q + 3q^2 + 2q^3 + q^4) x^4\\
		&& + (6 + 9q + 10q^2 + 9q^3 + 7q^4 + 3q^5 + q^6) x^2\\
		&& - (1 + 2q + 3q^2 + 3q^3 + 3q^4 + 2q^5 + q^6)\;.
	\end{array} 
\end{equation}

The $q$-Hermite polynomials reduce to Hermite polynomials for $q = 1$, i.e. $H_n(x|q=1) = He_n(x)$. Also, they reduce to Chebyshev polynomials for $q=0$, i.e. $H_n(x|q=0) = U_n(x/2)$, and they are defined by the following recurrence relation,
\be
2 x U_n(x) = U_{n-1}(x) + U_{n+1}(x);
U_{0}(x) = 1,\; U_{-1}(x) =0\;.
\ee
The $q$-normal distribution $f_{qN}(x|q)$ is given by \cite{KM2020},
\be
f_{qN}(x|q)  =  \dis \frac{1}{2\,\pi}\;\sqrt{\frac{1}{x_0^2-x^2}} \dis\prod_{i = 0}^{\infty}(1-q^{i+1})\;\l[(1+q^i)^2-q^i\;4\;\frac{x^2}{x_0^2}\r];
\label{eq:fqn}
\ee
where $x_0^2=4/(1-q)$ and $x$ is the standardized variable with zero centroid and unit variance. $f_{qN}(x|q)$ is defined over the interval $s(q)=(\frac{-2}{\sqrt{1-q}}, \frac{2}{\sqrt{1-q}})$. In the present study $q$ takes value between 0 and 1. Also it is plausible that, with forward shift operator, $D_q\; \{H_{n+1}(x|q)\} = \cac_{n,q} \; [n]_q\;H_{n}(x|q)$, and with backward shift  operator
$D_q\;\{ \can_{n,q}\, f_{qN} H_{n-1}(x|q)\} = - f_{qN} H_{n}(x|q)$. Here factors
$\cac_{n,q}$ and $\can_{n,q}$ depend on $n$ and $q$ only. Another important property is that the $q$-Hermite polynomials are orthogonal with respect to the $q$-normal weight function, $f_{qN}(x|q)$, over the range $s(q)$, i.e.
\begin{equation}
	\int_{s(q)} H_n(x|q)\; H_m(x|q)\; f_{qN}(x|q)\; dx = [n]_q! \;\delta_{mn}.
\end{equation}
Note that, $\int_{s(q)} f_{qN}(x|q)\;dx=1$. It can be easily seen that $f_{qN}(x|1) = \frac{1}{\sqrt{2\pi}} \; \exp(-x^2/2)$ and $f_{qN}(x|0) = \frac{1}{2\pi} \; \sqrt{4-x^2}$. Thus, $f_{qN}(x|q)$ interpolates between Gaussian to semicircle as $q$ changes from 1 to 0. Recently, the formula for $q$ as a function of ($m,N,k$) are obtained both for EGOE($k$) and BEGOE($k$) in \cite{Vyas2019} using their normalize fourth moment ($\mu_4=q+2$) and numerical tests confirm that $f_{qN}(x|q)$ describes the eigenvalue density very well for all $k$ values \cite{KC2018,Vyas2019,rao21}.

\section{Average-Fluctuation Separation in EGOE($k$) and BEGOE($k$)}
\label{sec:4}

\begin{figure}[!th]
	\begin{center}
		\begin{tabular}{cccc}
	\multicolumn{2}{c}{\includegraphics[width=0.45\linewidth]{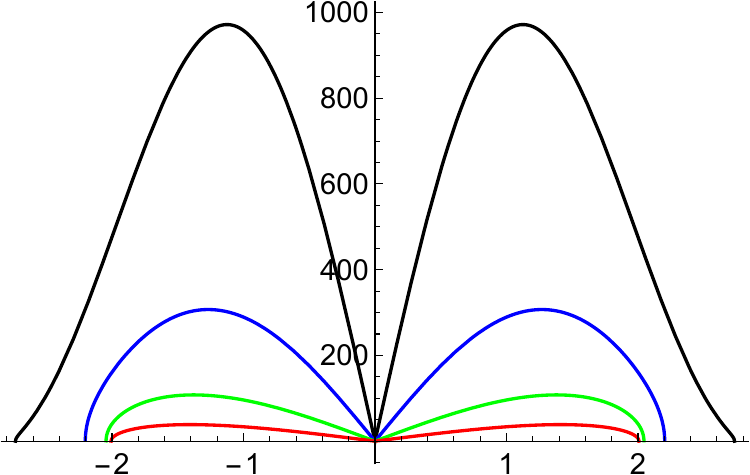}}&\multicolumn{2}{c}{\includegraphics[width=0.45\linewidth]{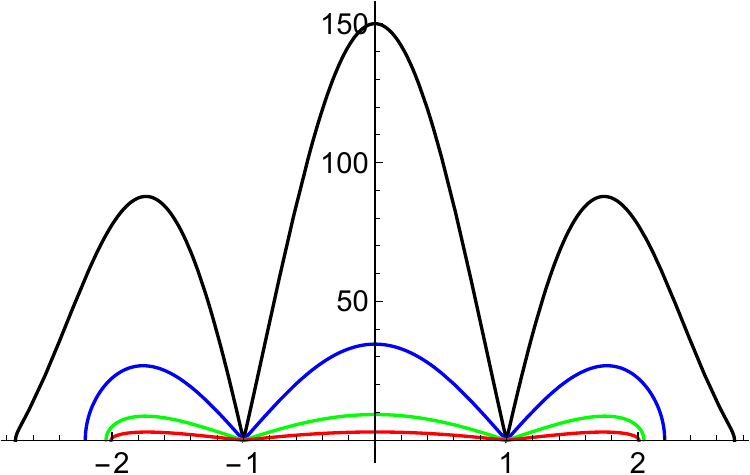}}\\
	(a)&$\hat{E}$&(b)&$\hat{E}$\\\\
	\multicolumn{2}{c}{\includegraphics[width=0.45\linewidth]{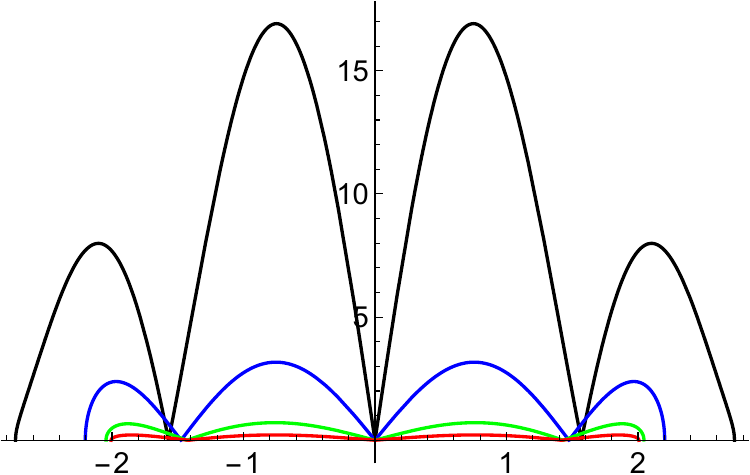}}&\multicolumn{2}{c}{\includegraphics[width=0.45\linewidth]{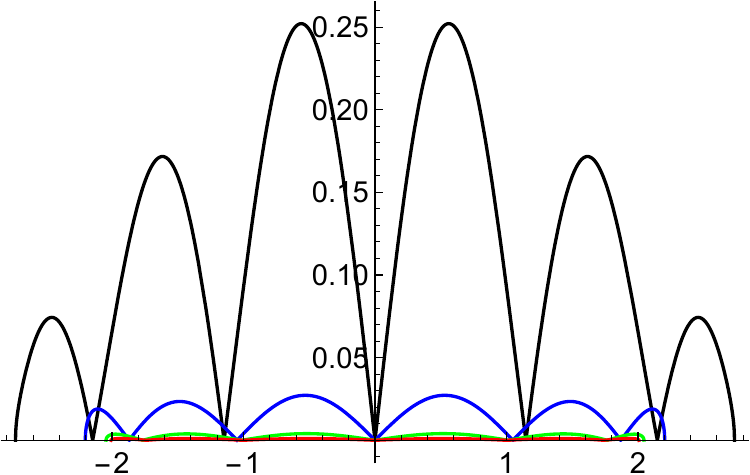}}\\
	(c)&$\hat{E}$&(d)&$\hat{E}$
\end{tabular}
	\end{center}
	\caption{Scaled widths of the level motion as a function of normalized energy given by Eq.~\eqref{eq:fnmd} for $m=10$ fermions in $N=20$ sp states. Results are shown for the excitation mode (a) $n=2$, (b) $n=3$, (c) $n=4$ and (d) $n=6$. In each panel, black curve for body rank $k=2$($q=0.465$), blue curves for $k=3$ ($q=0.176$), green curves for $k=4$ ($q=0.044$) and red for $k=5$ ($q=0.007$). The $q$ values given in the bracket are obtained using Eq.(12) of Ref.~\cite{Vyas2019}.}
	\label{fig:1}
\end{figure}

\begin{figure}[!th]
	\begin{center}
		\begin{tabular}{cccc}
	\multicolumn{2}{c}{\includegraphics[width=0.45\linewidth]{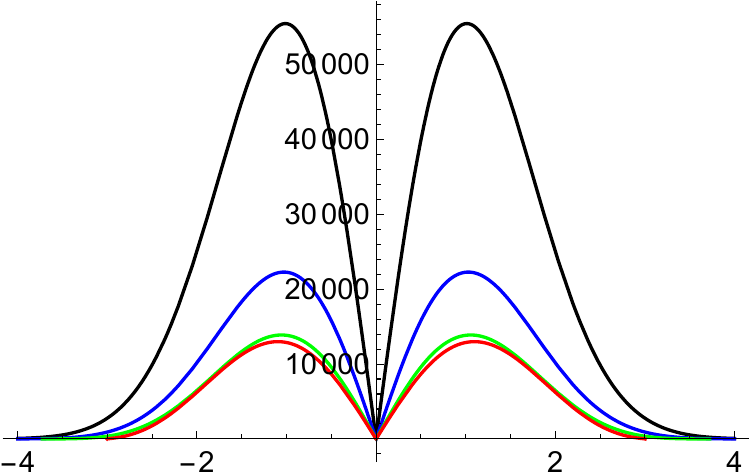}}&\multicolumn{2}{c}{\includegraphics[width=0.45\linewidth]{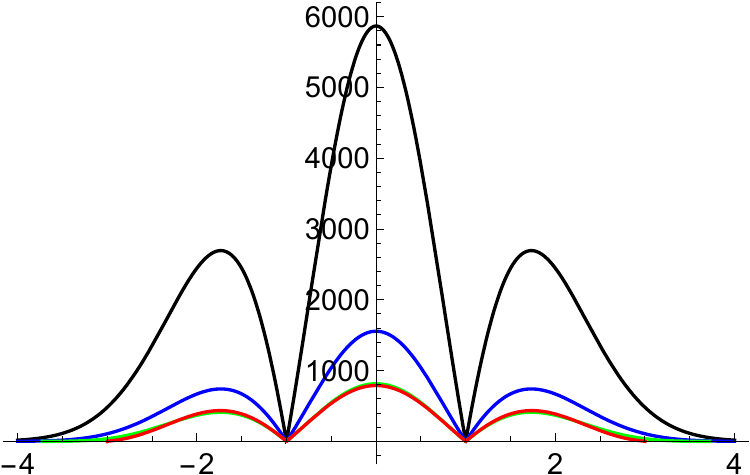}}\\
	(a)&$\hat{E}$&(b)&$\hat{E}$\\\\
	\multicolumn{2}{c}{\includegraphics[width=0.45\linewidth]{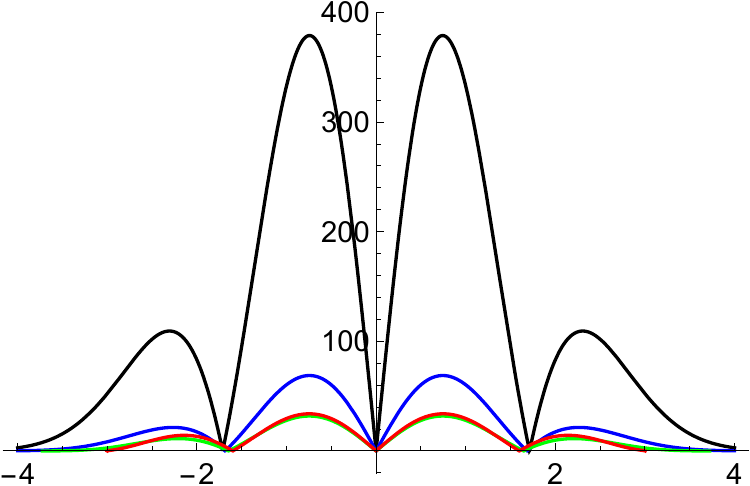}}&\multicolumn{2}{c}{\includegraphics[width=0.45\linewidth]{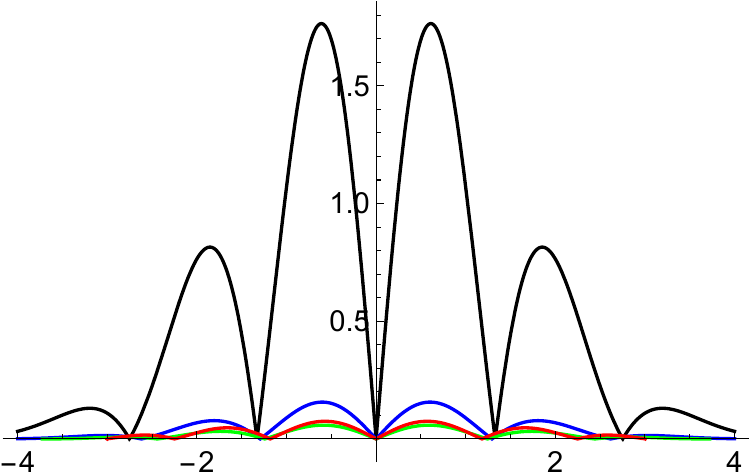}}\\
	(c)&$\hat{E}$&(d)&$\hat{E}$
\end{tabular}
	\end{center}
	\caption{Same as Figure \ref{fig:1} but for boson system with $m=20$ bosons in $N=10$ sp states, using Eq.~\eqref{eq:bnmd}. Results are shown for  the excitation mode (a) $n=2$, (b) $n=3$, (c) $n=4$ and (d) $n=6$. In each panel, the black curve for $k=2$($q=0.932$), blue for $k=3$ ($q=0.84$), green for $k=4$ ($q=0.712$) and red for $k=5$ ($q=0.556$). The $q$ values given in the bracket are obtained using Eq.(13) of Ref.~\cite{Vyas2019}.}
	\label{fig:2}
\end{figure}

Let us consider a normalized state density $\rho(E)$ and its smooth form given by $\overline{\rho}(E)$. With orthogonal  polynomials $P_{n}(E)$ defined by smooth density $\overline{\rho}(E)$,
such that $\int_{-\infty}^{\infty} dE \; \overline{\rho}(E)\; P_{n}\; P_{n'} = \delta_{n,n'}$ \cite{kota-book,Szego2003}, one can expand $\rho(E)$ using the polynomials $P_n(E)$ as,
\begin{equation}
\rho(E)=\overline{\rho}(E) \l\{ 1+ \sum_{n=1} S_{n} P_{n}(E) \r\}\;.
\label{eq:gc}
\end{equation}
Now, EGOE($k$) for fermions in the dilute limit  (and also BEGOE($k$) for bosons in the dense limit), the ensemble averaged state density changes from Gaussian to semi-circle form as $k$ changes from 1 to $m$, and can be represented by $q$-normal distribution \cite{KC2018,Vyas2019,rao21} given by,
\be
\begin{array}{rcl}
\rho_{q}(E)\, dE  &=&  \dis \frac{1}{4\pi \sigma}\,\sqrt{\frac{1-q}{1-\l( \frac{E-\epsilon}{E_0}\r)^2}} \dis \prod_{i = 0}^{\infty} (1-q^{i+1}) \l[(1+q^i)^2-q^i\;4\;\l( \frac{E-\epsilon}{E_0}\r)^2 \r] dE;\\\\

&&\dis E_0^2=\frac{4\sigma^2}{1-q},\;\; \dis \epsilon - \frac{2\,\sigma}{\sqrt{1-q}} \leq E \leq \epsilon + \frac{2\,\sigma}{\sqrt{1-q}}.

\end{array}
\label{eq:fqne}
\ee
Here, $\epsilon$ and $\sigma^2$ are centroid and variance of $\rho_q$. One can expand $\rho(E)$ in terms of the smooth density $\overline{\rho}(E)=\rho_q(E)$, using the $q$-Hermite polynomials $H_{n}(\hat{E}|q)$ as,
\begin{equation}
	\rho(E) = \rho_q(E) \l\{ 1+ \sum_{n} ([n]_q!)^{-1} \; S_{n} \; H_{n}(\hat{E}|q) \r\}\;.
	\label{eq:gcc}
\end{equation}
In the above equation centroid and variance of $\rho$ is same as $\rho_q$ and $S_{n}$ can be connected to the higher moments of $\rho$. As $S_n$'s vary from  member to member of the ensemble, they can be taken random variables with centroid zero, i.e. $\overline{S_n}=0$. It is important to note that $q = 1$ in Eq.~\eqref{eq:fqn} gives the Gaussian, the Eq.~\eqref{eq:gcc} leads to the Gram-Charlier expansion starting with Gaussian that takes into account first two moments, centroid and variance. While for $q \in (0,1)$, $f_{qN}$ takes into account the second and fourth moment (odd moments are zero). In  Eq.~\eqref{eq:gcc}, each $n$ term represents a mode of the excitation and wavelength of the mode is inversely proportional to $[n]_q$. Also, both EGOE($k$) and BEGOE($k$) with $k=m$, exactly give GOE, $H_{n}(\hat{E}|q=0)$ reduces to the Chebyshev polynomials in Eq.\eqref{eq:gcc}. With exact distribution function taken as $F(E)= d(N,m) \int\limits_{ - 2/\sqrt{1-q} }^E {f_{qN} (x|q) dx}$, the difference between
energy level $E$ and its smooth part $\overline{E}$, with respect to the ensemble, gives level motion in terms of local mean spacing ($\overline {D(E)}$) and is given as
${\delta E}=E-\overline{E}= \; \left[ F(E) - \overline {F(E)}\right] \overline {D(E)}$. The variance of level motion, $\overline {{{( {\delta E} )^2 }}}/ {{\overline{D(E)} ^2 }}$, after including centroid and variance fluctuations, is given by,

\begin{eqnarray}
\dis	\frac{\overline {( {\delta E} )^2}}{{\overline{D} ^2}} &=& \overline{\left[ F(E) - \overline
		{F(E)}\right]^2} \nonumber \\ &=& d^2 \;\sigma^2\; f_{qN}(E|q)^2 \\ &&\times \left\{ {\sum\limits_{n \geqslant
			1} {\can_{n,q}^2 \left( {[n]_q !} \right)^{-2} \overline {\left( {S_n ^2
				} \right)} \;\left[ {H_{n  - 1} \left( {\hat E}|q \right)}
			\right]^2 } } \right\}.\nonumber
\label{var2}
\end{eqnarray}
Hence, $\overline {\left( {S_n^2} \right)}$ are needed for EGOE($k$) and BEGOE($k$) and they are related to the co-variances $\Sigma_{p,q} = \overline{
{\langle H^p\rangle}{\langle H^q \rangle }}  - \overline {\langle H^p \rangle}\,\overline {\langle {H^q} \rangle}$. The variance of $m$-fermion is  $\sigma_{EGOE}^{2}(m)=\binom{m}{k} {\l[\binom{N-m+k}{k}+1 \r]}$ and that of $m$-boson  is $\sigma_{BEGOE}^{2}(m)=\binom{m}{k} \binom{N+m-1}{k}$. With normalization $\sigma^2(k)=1$, the formula for $\overline {\left( {S_n ^2} \right)}$, for EGOE($k$) in strict $N \rightarrow \infty$ limit can be given by \cite{kota-book},
\begin{equation}
	\overline {\left( {S_n  }\right)^2 }  = 2 n {\binom{m}{k}}^{2-n } {\binom{N}{k}}^{-2}.
\end{equation}
Similarly, in strict $m \rightarrow \infty$ limit for BEGOE($k$), it is plausible that \cite{kota-book},
\begin{equation}\label{ss}
	\overline {\left( {S_n ^2 } \right)}  = \dis 2 n\; {\binom{N}{k}}^{-n }.
\end{equation}
The formulas for $\overline {\left( {S_n ^2 } \right)}$ reduce to $k=2$ case derived using binary correlation approximation in \cite{Brody81,kota-book,MF1975,PDPK2000}. With $k$ body interactions, the result for level motion in the fermion systems is,
\be
\begin{array}{rcl}
\dis	\frac{\overline {( {\delta E} )^2}}{{\overline{D} ^2}} &=& \dis {N \choose m}^2  { m \choose k}^{2} {\binom{N}{k}}^{-2} \; \rho_q(E)^2 \\ && \dis\times \left\{ {\sum\limits_{n
			\geqslant 1} {\left( {[n_q] !} \right)^{ - 2} 2 \,n\, {m \choose  k}^{2 - n } \left[ {H_{n  - 1} (\hat E |q )} \right]^2 } } \right\}\;,
\end{array}
\label{eq:fnmd}
\ee
and in the boson systems is,
\be
\begin{array}{rcl}
\dis	\frac{\overline {( {\delta E} )^2}}{{\overline{D} ^2}} & = & \dis {N+m-1 \choose m}^2 {m \choose k}^2\; {N \choose k}^{-2}\; \rho_q(E)^2
\\ && \dis\times  \left\{ {\sum\limits_{n \geqslant 1}
		{\left( {[n]_q !} \right)^{ - 2} 2\, n \, {N \choose k}^{-n} \left[ {H_{n - 1}
				\left( {\hat E|q} \right)} \right]^2 } } \right\}.
\end{array}
\label{eq:bnmd}
\ee
The width of level motion for different mode of excitation $n$, scaled by factor $\can_{n,q}^2$ for various values of $k$, are obtained via  Eq.~\eqref{eq:fnmd} for fermions and Eq.~\eqref{eq:bnmd} for bosons and results are shown in Figures \ref{fig:1} and \ref{fig:2}, respectively. The results are obtained for the following examples: (i) $m=10$ fermions in $N=20$ sp states with dimensionality of the space $d=184756$ and (ii) $m=20$ bosons in $N=10$ sp states with dimensionality of the space $d=10015005$. The $q$ values here are obtained using Eq.~(12) for fermions and Eq.~(13) for bosons of \cite{Vyas2019}. The formulas in Eqs.~(12)-(15) are good only for $k << m$ and they do not apply as $k$ goes towards $m$. For $k=m$ (GOE), the normal mode decomposition is given by Eq.~(4.37) in Brody et al \cite{Brody81}. The point of Eqs.~(14) and (15) is that these expansions are quite different from the expansion for GOE ($k=m$) as argued in \cite{Brody81} (see the discussion after Eq.~(4.37) in \cite{Brody81}). One can draw following conclusions from the results shown in Figures~\ref{fig:1} and \ref{fig:2}: (i) The intensity of fluctuations rapidly goes to zero as the mode of excitation $n$ and rank of the interaction $k$  increases. (ii) The rate of fluctuations decreases faster with respect to $n$ indicating sharp separation and it also decreases with increasing  $k$ value. This is due to ${m \choose k}$ terms in Eq.\eqref{eq:fnmd} for fermions and ${N \choose k}$ terms in Eq.\eqref{eq:bnmd} for bosons. (iii) The wavelength of excitation mode decreases as both $n$ and $k$ increases. It is inversely proportional to $\sqrt{1-q}$ for a fixed mode of excitation. (iv) For smaller value of $k$, smooth part of the state density can be defined using only a few long wavelength modes. (v) With higher rank of interaction, the intensity of excitation mode decreases considerably and therefore the number of modes require to define averages effectively reduced. (v) The fluctuations set in faster for higher rank of interactions $k$. In the next section, this behavior will be verified numerically using EGOE($k$) and BEGOE($k$) examples.

\section{Numerical Results for fermionic EGOE($k$) and bosonic BEGOE($k$)}
\label{sec:5}

In order to analyze the separation between average and fluctuation part in the spectra of EGOE($k$) and BEGOE($k$) for a given body rank of the interaction $k$, following examples are considered: (i) EGOE($k$) for $m = 6$ fermions in $N = 12$ sp states with $H(m)$ matrix of dimension $d=924$. (ii) BEGOE($k$) for $m = 10$ bosons in $N = 5$ sp states with $H(m)$ matrix of dimension $d=1001$. It is important to note that the dimensionality of $H(m)$ increases rapidly with the number of particles $m$ and/or sp states $N$ and the numerical calculations will be prohibitive for large $m$ and $N$ values. Therefore, the selection of systems in the present analysis are based on limitations imposed by computational facilities. However, it is clear that the fermionic EGOE($k$) with ($m=6$, $N=12$) example exhibits the properties of dilute limit \cite{Lec2008,Vyas2019,rao21a} and the bosonic BEGOE($k$) with ($m=10$, $N=5$) example exhibits the properties of dense limit \cite{PDPK2000,Lec2008,Vyas2019,rao21,DCP13,rao21a}. An ensemble of 50 members each is used for given $k$ value in both the examples and corresponding ensemble averaged skewness ($\gamma_1$) and excess ($\gamma_2$) parameters are given in Table \ref{table:1}.
\begin{table}[!htb]
	\centering
	\caption{The ensemble averaged skewness ($\gamma_1$) and excess ($\gamma_2$) parameters for EGOE($k$) and BEGOE($k$) examples.}
	\begin{center}
	\begin{tabular}{|ccc|ccc|}\hline\hline
	\multicolumn 3 {|c|}{EGOE($k$):$m=6,N=12$} &
	\multicolumn 3 {|c|}{BEGOE($k$):$m=10,N=5$} \\ \hline\hline
	 $k$& $\gamma_1$ & $\gamma_2$ &$k$& $\gamma_1$ & $\gamma_2$ \\ \hline
	 $2$& 0.0023 &-0.7172 &$2$& -0.0025&-0.1463 \\
	 $3$& -0.0002 & -0.9422 &$3$&-0.0125 &-0.3083\\
	 $4$& 0.0001& -0.9945 &$4$& 0.0024 &-0.5834 \\
	 $5$& 0.0001 & -0.9980  &$5$& 0.0013& -0.8205 \\
	 $6$& -0.0003 &-0.9991&$6$& 0.0005 &-0.9504 \\
	 &&&$7$& 0.0000 &-0.9909 \\
	 &&&$8$& 0.0000 & -0.9950 \\
	 &&&$9$& 0.0000 & -0.9984 \\
	 &&&$10$& 0.0000 & -0.9995\\
 \hline \hline
\end{tabular}
\end{center}\label{table:1}
\end{table}

\subsection{Normal mode decomposition of the spectra}

From the eigenvalues obtained via matrix diagonalization of $H(m)$, one can compute the exact distribution function $F(E)$. While the smooth part, $\overline{F(E)}$, for a given order $n_0$, follows  from the smooth form of state density,
\begin{equation}
\overline{\rho(E)}= \rho_q(E)\;\l\{ 1+ \sum_{n \geq
	3}^{n_0} \l( [n_q] !\r)^{-1}\; S_{n}\;
H_{n}(\hat{E}|q)\r\}\label{gc1}\;.
\end{equation}
The level motion $\Delta(E) = F(E)- \overline{F(E)}$ with $\overline{F(E)}=F_q(E)$ is computed following the procedure described in \cite{PDPK2000,Lab90}. The $q$ value is obtained separately for each member using normalized fourth moment of the spectra. As the energy spectrum is discrete, the $S_{n}$'s in Eq.\eqref{gc1}, for a given order $n_0$, are determined by minimizing $\sum_{i}^{d}\{ \Delta(E_i)^2 \}$ with respect to the $S_{n}$'s. With these optimized $S_{n}$'s give level motion $\Delta(E)$ vs $E$ for the order $n_0$. Here, the ensemble average, for given $k$ and $n_0$, is carried out after the spectra of each member of the ensemble is first made zero centered and scaled to unit width.

\begin{figure}[!th]
	\begin{center}
		\includegraphics[width=\linewidth]{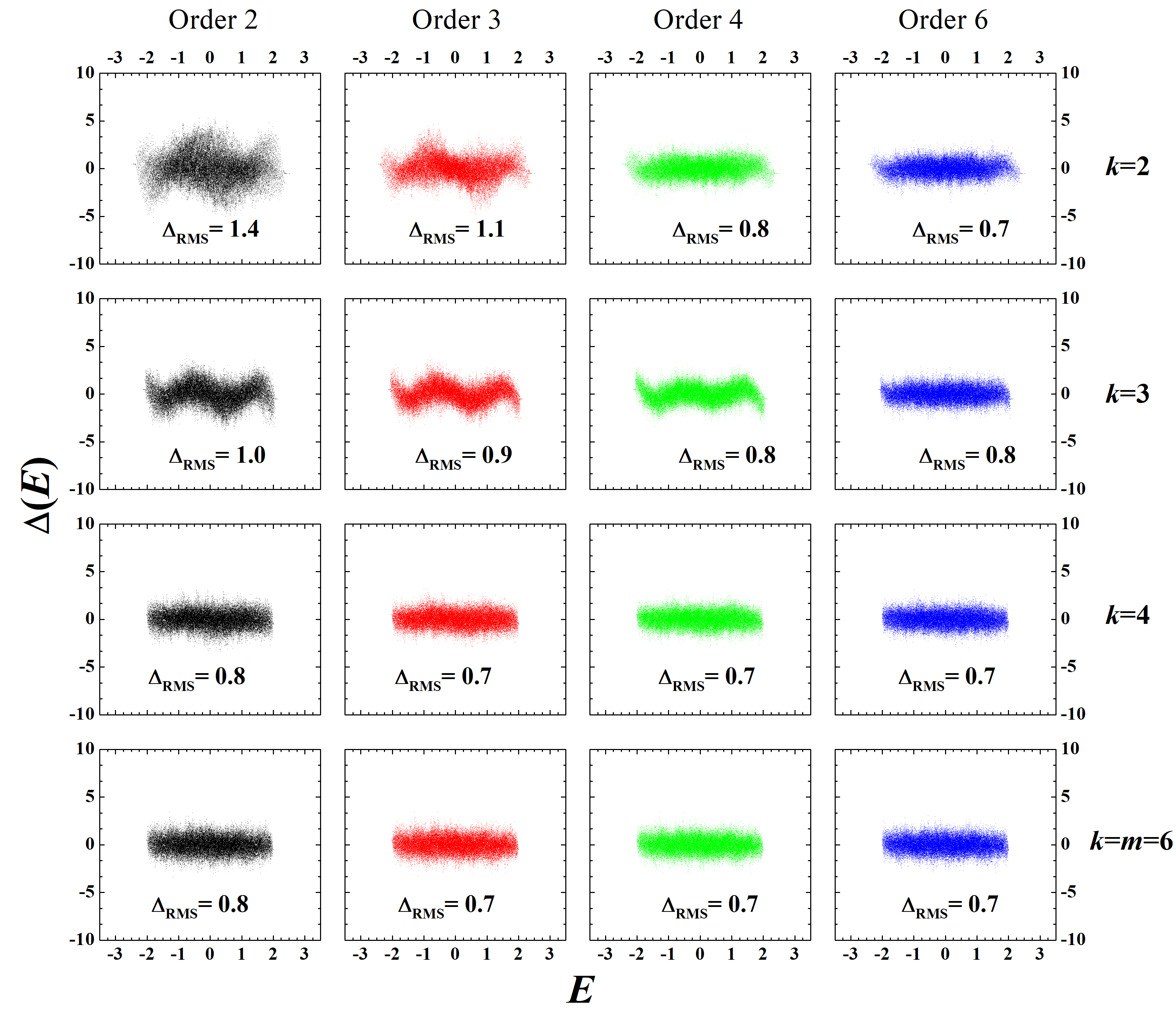}
	\end{center}
	\caption{Level motion $\Delta(\hat{E})$ vs. normalized energy $(E -\epsilon(k))/\sigma(k)$ are shown for the spectra of a 50 member EGOE($k$) generated using the system of 6 fermions in 12 sp states for different values of body rank $k$. The State-to-state deviation between the true distribution function and its smooth version with $\overline{\rho(E)}$ including optimized correction upto third order, fourth order and sixth order are shown for all members of the ensemble. First column is due to $\overline{F(E)}=F_q(E)$ (order 2).  Ensemble averaged $\Delta_{RMS}$ are also shown in the figure.}
	\label{fig:3}
\end{figure}

\begin{figure}[!th]
	\begin{center}
		\includegraphics[width=\linewidth]{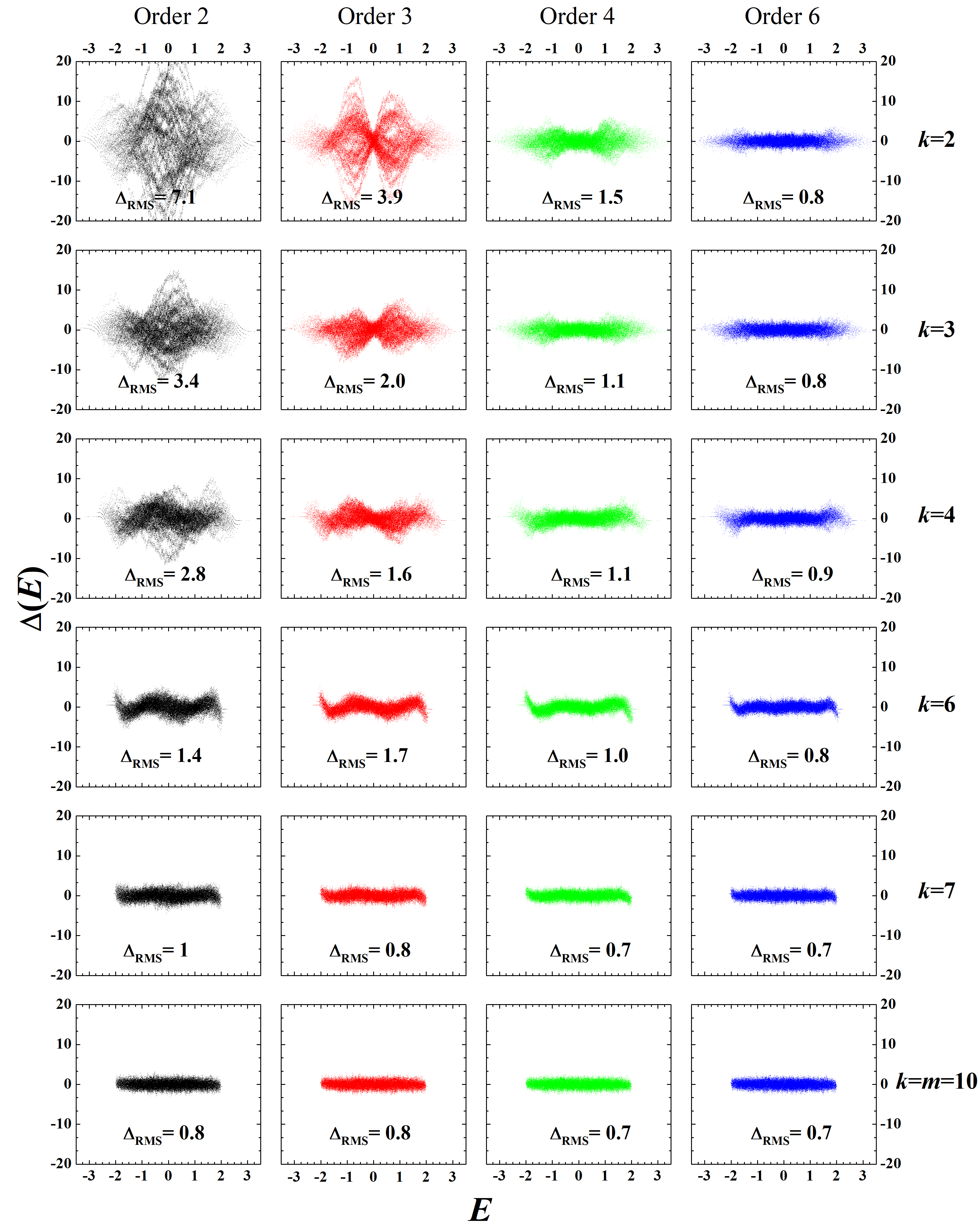}
	\end{center}
	\caption{Same as Figure \ref{fig:3} but for the spectra of a 50 member BEGOE($k$) generated using the system of 10 bosons in 5 sp states for different values of body rank $k$. Ensemble averaged $\Delta_{RMS}$ are also shown in the figure.}
	\label{fig:4}
\end{figure}

Figure~\ref{fig:3} shows the results for the normal-mode decomposition of the spectra for EGOE($k$) generated by 6 fermions in 12 sp states for different values of the rank of interaction $k$. The first column in Figure~\ref{fig:3} corresponds to $[\Delta(E) = F(E)- F_q(E)]$ vs $\hat{E}$. The results for $\Delta(E)$ with $\overline{\rho(E)}$ optimized to orders $n_0=3$, 4 and 6 are also shown. In the ensemble calculations for given $k$, level motion for each order ($n_0$) is obtained separately for each member and the results in the figure are shown for all the members. Similarly, Figure~\ref{fig:4} shows the results for the normal-mode decomposition of spectra for BEGOE($k$) generated by 10 bosons in 5 sp energies for different $k$ values. The values for root mean square deviations $\Delta_{RMS}$ in $\Delta(E)$ for EGOE($k$) and BEGOE($k$) examples are also shown in the corresponding panel in Figures~\ref{fig:3} and \ref{fig:4}. It is clearly seen, for both EGOE($k$) and BEGOE($k$), that the $\Delta_{RMS}$ falls sharply at first with increasing order and then varies slowly suggesting a clear cut dissociation between the smooth and fluctuating part of the distribution function. For fermionic EGOE($k$) example, for the rank of interactions $k<4$ and $n_0 \leq 3$, $\Delta_{RMS}$ value is larger than GOE value, which is $\sim 0.88$ at $\hat{E}=0$; the GOE formula for $\Delta_{RMS}^2=\ln{2 \;d(m,N)}/\pi^2$ \cite{Brody81}. By the time fourth order $n_0=4$ corrections are added to the asymptotic $q$-normal density, $\Delta_{RMS} \sim 0.8$ and which implies the onset of GOE fluctuations. Similarly, for BEGOE($k$) example, for the rank of interactions $k<6$, $\Delta_{RMS} \sim 1$ by the time fifth order corrections are added to $\overline{\rho(E)}$. With the sixth order corrections to $\overline{\rho(E)}$, $\Delta_{RMS}$ reaches close to the GOE result. These results are consistent with that of obtained for $k=2$ in \cite{PDPK2000,Lec2008}. Going further, with $k>4$ for fermion (and $k>6$ for bosons), the averaged density approaches semi-circle form and $q \rightarrow 0$. $\Delta_{RMS}$ comes close to GOE value with just $\overline{\rho(E)}=\rho_q(E)$.  The results here, clearly show distinction between the smooth and fluctuating part of the distribution function for all $k$ values. Also, the GOE fluctuations set in with $\overline{\rho(E)}$ including a few order of corrections as the rank $k$ increases.

In the past, using the analogy between energy spectrum and discrete time series, the spectral fluctuations were investigated in embedded ensembles using normalized periodogram analysis and it was shown that interacting quantum systems exhibit a clear cut average-fluctuation separation\cite{Lec2008}. It is important to note that the periodogram analysis was also emphasized in terms of $1/f$-noise signature of quantum chaos \cite{Rel1,Rel2,Rel3}. In the next section, we consider the method of a normalized periodogram to analyze EGOE($k$) and BEGOE($k$) spectra.

\subsection{Periodogram analysis}

The method of a normalized periodogram, introduced by Lomb and Scargle \cite{Lomb-Scargle}, was initially used for analyzing astronomical data and it is very efficient in testing and finding the presence of weak periodic signals in unevenly spaced data. The Lomb-Scargle statistic estimates mainly the two-parameters. First is $f_p$, the frequency at which the maximum value of periodogram occurs and second is the parameter $\Lambda$ (in percentage) measuring the relevance of the signal at $f_p$ against the presence of random noise \cite{Lec2008}. Therefore, in terms of the parameter $\Lambda$, the periodogram analysis provides an accurate quantitative measure giving the extent to which the separation between smooth and fluctuating parts exist.

\begin{figure}[!th]
	\begin{center}
		\includegraphics[width=\linewidth]{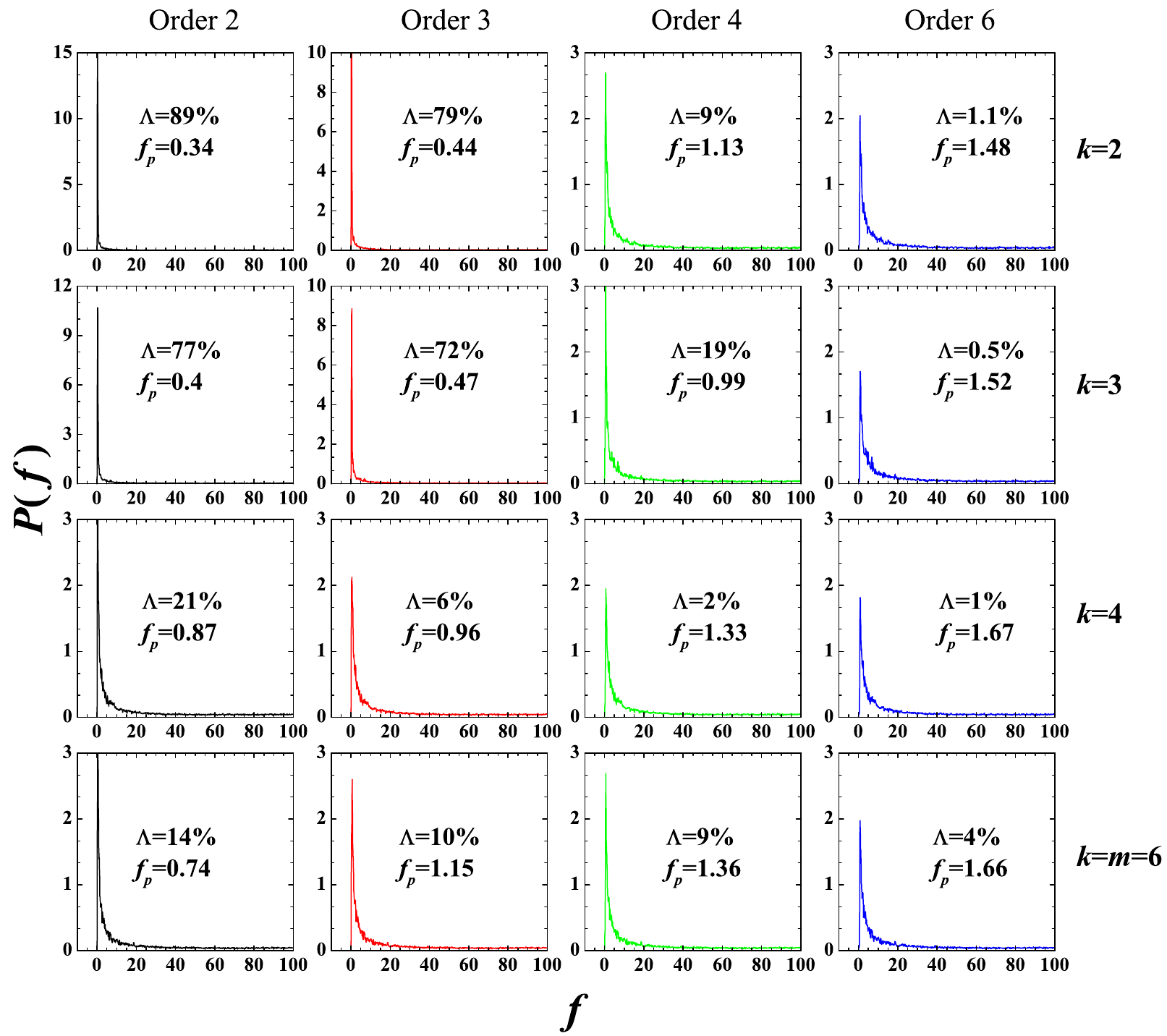}
	\end{center}
	\caption{Periodogram $P(f)$ vs $f$ for the corresponding deviation data shown in Figure \ref{fig:3} for EGOE($k$) ensemble. Note that $f$ and $P(f)$ are unit less. The values of parameters, $\Lambda$ and $f_p$ are shown in the figure. See text for details.}
	\label{fig:5}
\end{figure}

\begin{figure}[!th]
	\begin{center}
		\includegraphics[width=\linewidth]{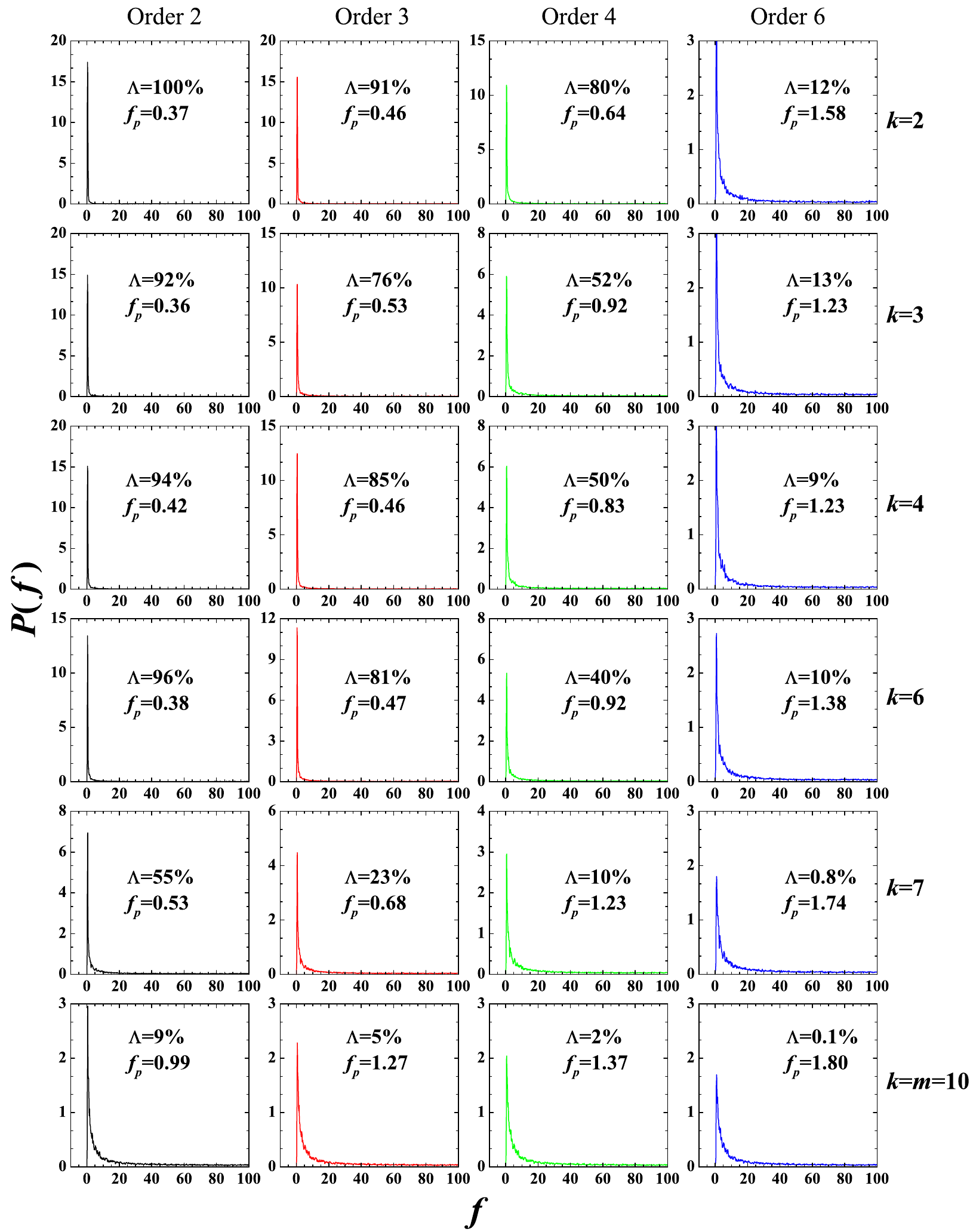}
	\end{center}
	\caption{Periodogram $P(f)$ vs $f$ for the corresponding deviation data shown in Figure \ref{fig:4} for BEGOE($k$) example. See Figure\ref{fig:5} and text for details.}
	\label{fig:6}
\end{figure}

Following the procedure given in \cite{Lec2008}, the power spectra are obtained using the deviations, between the smooth part and actual part of the distribution functions, obtained in the previous section. For a given order $n_0$ and body rank $k$, the periodogram $P(f)$ vs $f$ is obtained, for each $\Delta(E)$ vs $E$ data shown in Figure~\ref{fig:3} for EGOE($k$) and Figure~\ref{fig:4} for BEGOE($k$), and the corresponding results are shown in Figures~\ref{fig:5} and \ref{fig:6}, respectively. The parameters $\Lambda$ and $f_p$ are calculated for each member of the ensemble separately using middle 90\% of the spectrum and then the ensemble averaged periodogram is obtained. The ensemble averaged values of the parameters, $\Lambda$ and $f_p$, are shown in Figures~\ref{fig:5} and \ref{fig:6} for EGOE($k$) and BEGOE($k$), respectively. For EGOE($k$) example with $k=$ 2 and 3, $\Lambda$ value is found to be greater than 70\% for 3rd order corrections included in the smooth density indicating presence of signal at this stage. The $\Lambda$ value reduces to less than 15\% by the time 4'th order correction is added. This implies the absence of a signal and indicating the presence of noise only. Hence, the onset of fluctuations occurs with 4th order correction added to the smooth density for $k \le 3$. Further, for $k\ge4$, $\Lambda$ value is found to be less than 15\% with just second order corrections in the smooth density and for higher order $\Lambda$ values further reduced indicating onset of fluctuations occur with just second order corrections.

Similarly, for BEGOE($k$) example, the $\Lambda$ values become small only after 6th order corrections
are taken into account for $2\le k\le 6$. Thus, there is average-fluctuation separation in energy
levels with averages determined by 6th order corrections for $k < 7$. While for $k=7$, the $\Lambda$ values reduces to 10\%  with fourth order correction. Further, for $k > 7$, the $\Lambda$ values become
very small with just 2nd order correction indicating onset of fluctuations with just second order corrections. The results here are consistent with previously obtained for $k = 2$\cite{PDPK2000,Lec2008}. Thus, there is average-fluctuation separation in energy levels with averages determined by 2-6 order corrections to the asymptotic density represented by $f_{qN}(x|q)$ for both, EGOE($k$) and BEGOE($k$). Also, the periodogram analysis, with $\Lambda$ parameter, clearly shows that the GOE fluctuations set in faster as the rank $k$ of the interaction increases. For both the examples, the power spectrum results are in good agreement with that of the normal mode decomposition analysis carried out in the previous section.

Now, using the unfolded spectra, NNSD and the Dyson-Mehta $\Delta_3$ statistic \cite{Dyson} will be studied to confirm that the fluctuations are of GOE type. There exists a wide variety of quantum systems which demonstrate GOE fluctuations \cite{Haake2010,Mehta}.

\subsection{NNSD and $\Delta_3$ statistic}

\begin{figure}[!th]
	\begin{center}
		\begin{tabular}{cc}
		(a)&	 \includegraphics[width=0.55\linewidth]{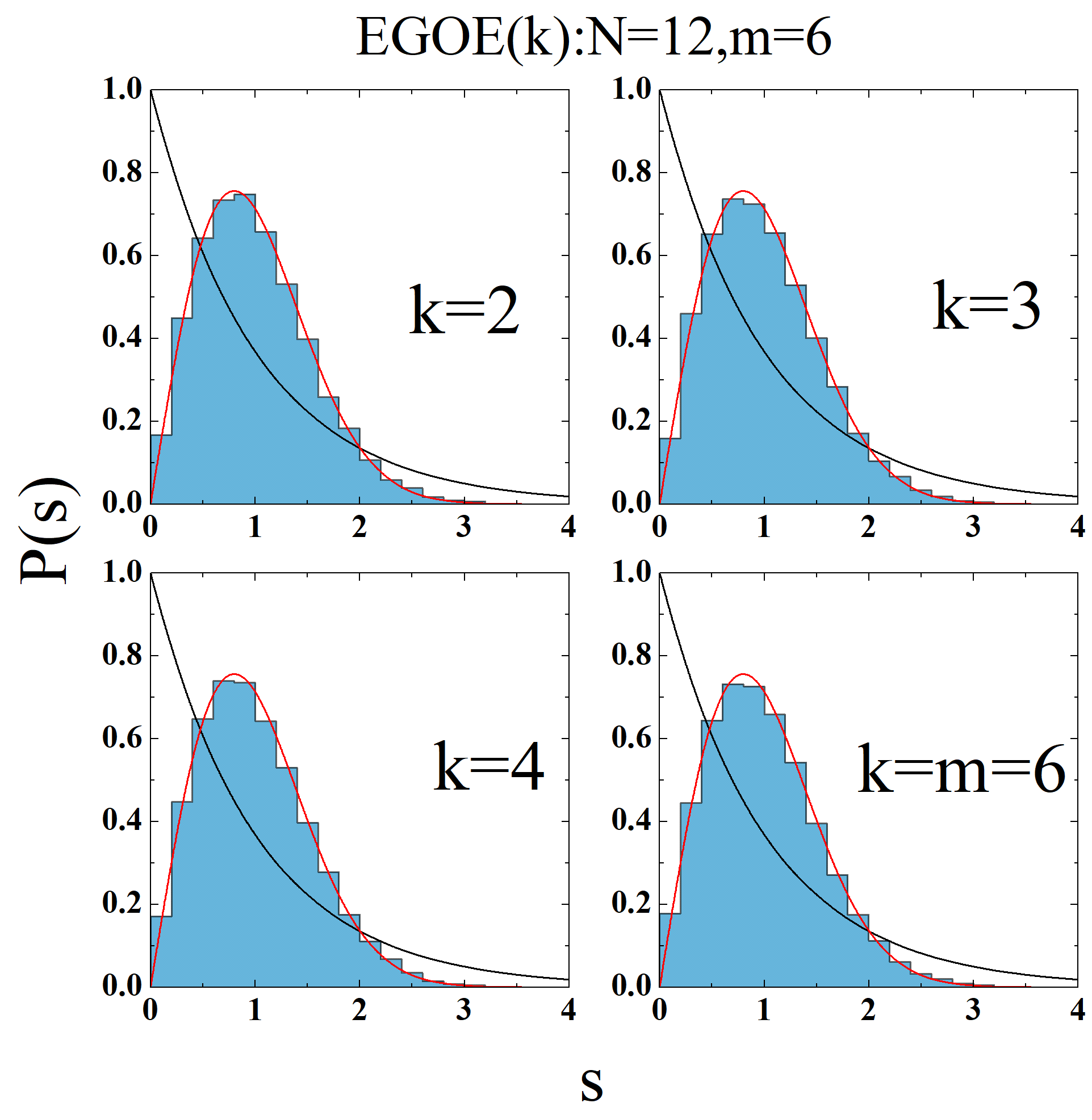}\\
		(b)&\includegraphics[width=0.8\linewidth]{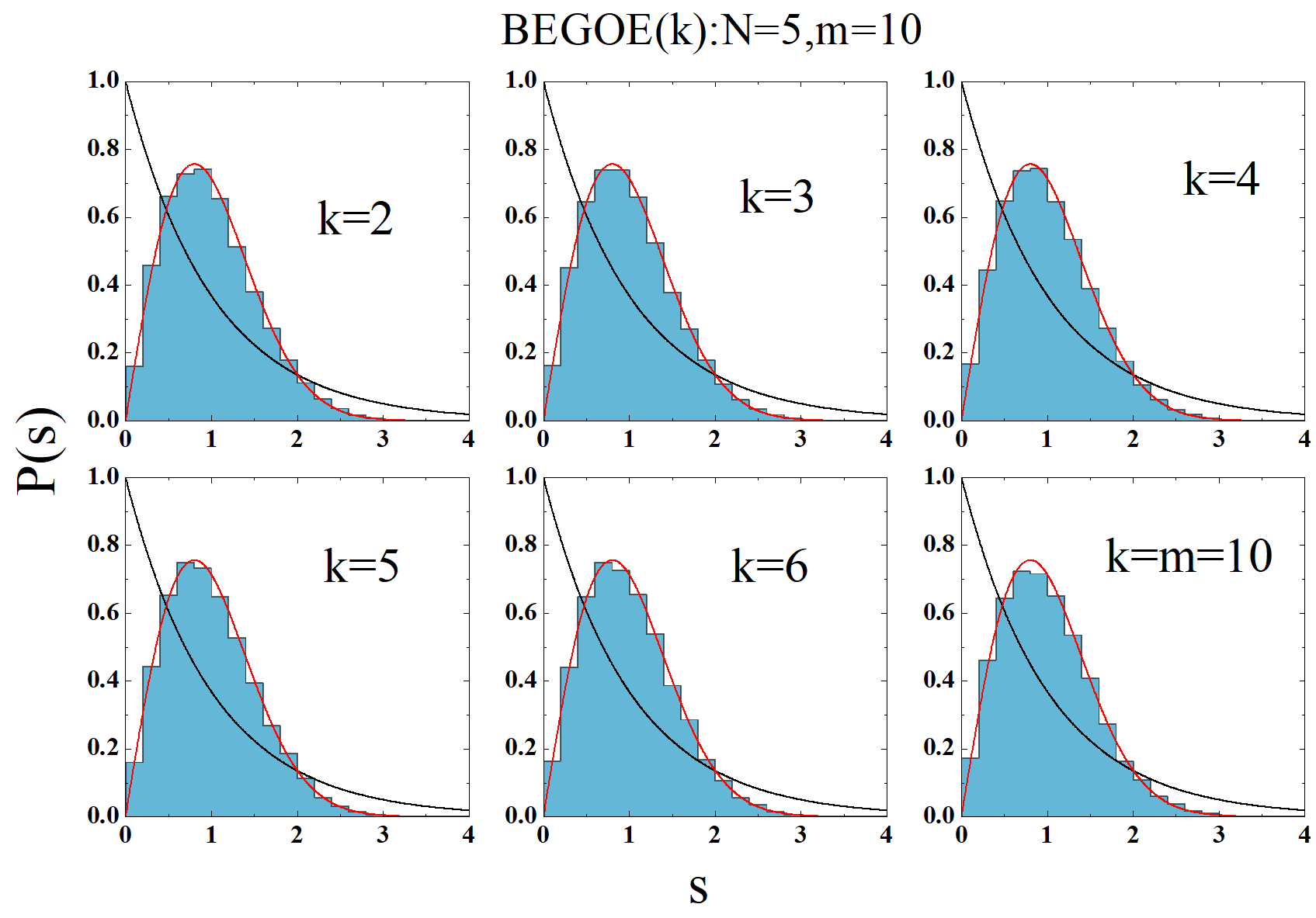}
		\end{tabular}
	\end{center}
	\caption{
Ensemble averaged NNSD histograms for a EGOE($k$) example ($m = 6;N = 12$) in (a)
and for a BEGOE($k$) example ($m = 10;N = 5$) in (b). Results are shown for various values of
body rank $k$. Also Poisson and GOE  predictions are superimposed. See text for further details.}
	\label{fig:7}
\end{figure}

\begin{figure}[!th]
	\begin{center}
		\begin{tabular}{cc}
			(a)&	 \includegraphics[width=0.55\linewidth]{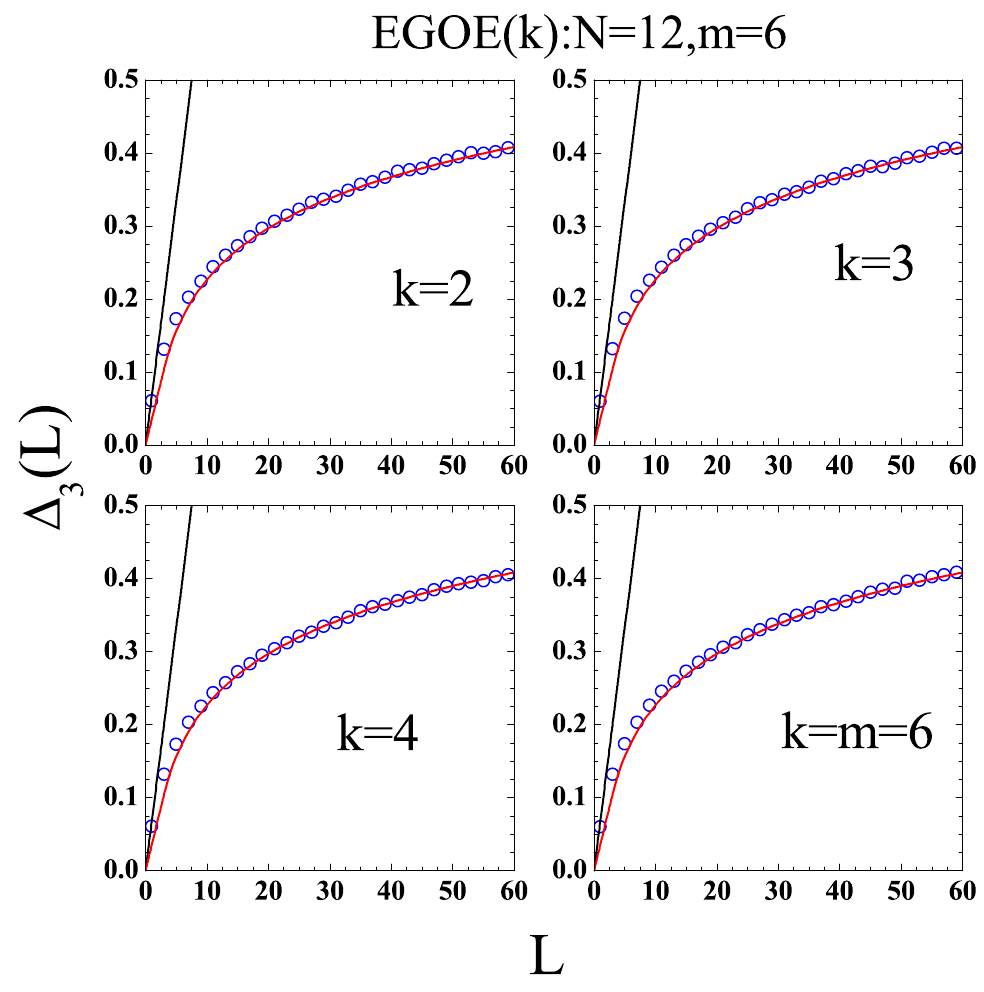}\\
			(b)&\includegraphics[width=0.8\linewidth]{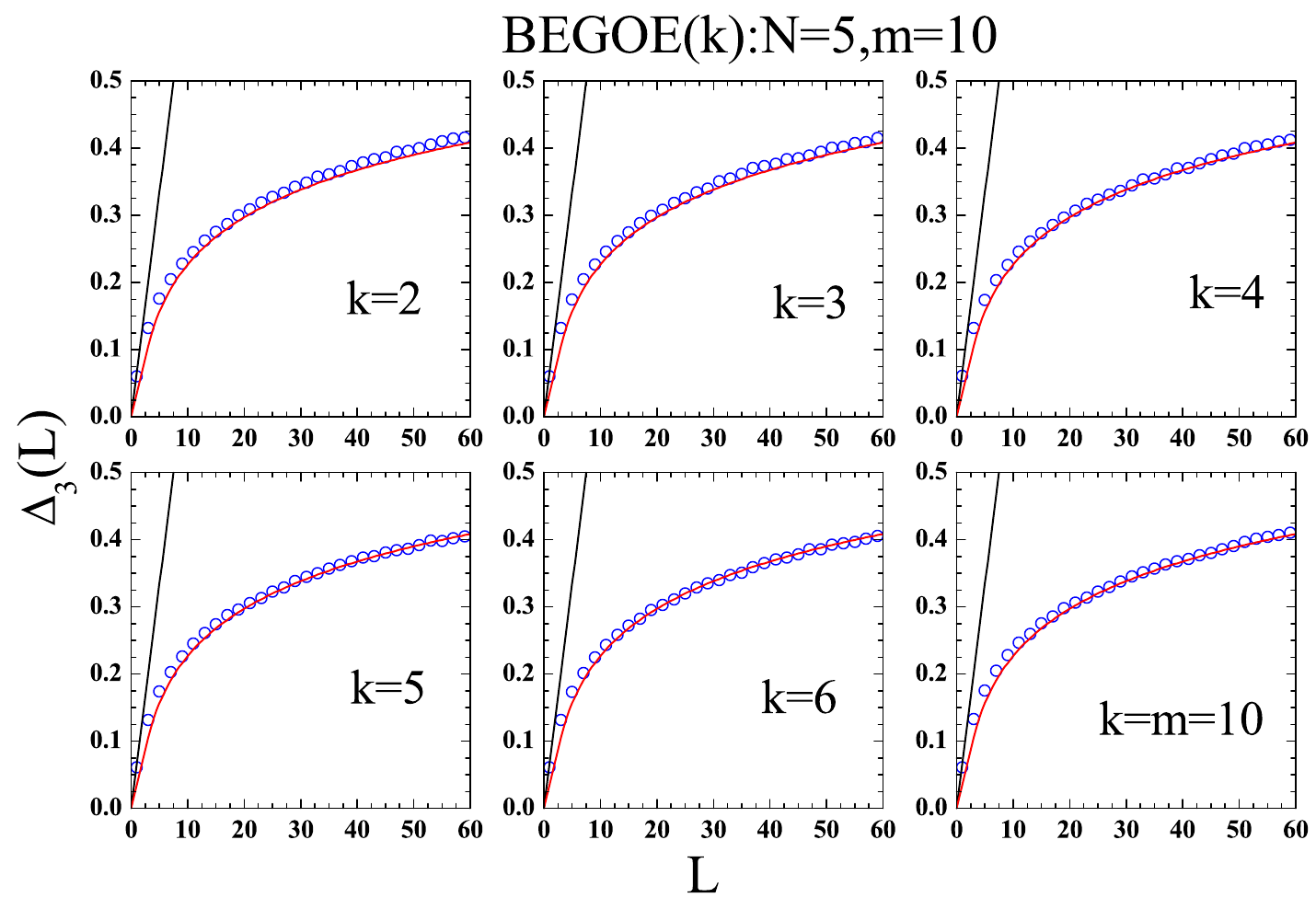}
		\end{tabular}
	\end{center}
\caption{$\Delta_3(L)$ vs. $L$ for EGOE($k$) example ($m = 6;N = 12$) in (a) and for a BEGOE($k$) example ($m = 10;N = 5$) in (b). Results are shown for various values of body rank $k$. Poisson and GOE predictions are also superimposed. See text for further details.}
\label{fig:8}
\end{figure}

In order to confirm the GOE limit, the NNSD and the $\Delta_3(L)$ statistic are constructed, for each $k$ value, using eigenvalues of EGOE($k$) and BEGOE($k$) examples considered above. NNSD and $\Delta_3$ are  constructed for various values of $k$ changing from 2 to $m$. The results for NNSD and $\Delta_3(L)$ are shown in Figures~\ref{fig:7} and \ref{fig:8} respectively. In these calculations, spectrum of each member is unfolded separately so that the average spacing is unity and then the ensemble averaged NNSD is constructed using the middle 90\% of the spectrum.
For EGOE($k$) example, the unfolding is done using $\overline{\rho(E)}$ with $n_0=4$ for $k<4$, while for $k>4$ $\overline{\rho(E)}=\rho_q(E)$ is taken. Similarly for BEGOE($k$) example, $\overline{\rho(E)}$ with $n_0=6$ is chosen for $k \le 7$ and $\overline{\rho(E)}=\rho_q(E)$ for $k>7$. The NNSD results are superimposed with the Poisson and Wigner forms. For $k=m$ case, both EGOE($k$) and BEGOE($k$) give exactly GOE. Therefore, the spectral unfolding is done with optimized second order corrections to the smoothed density for both fermionic as well as bosonic examples. It is clear from the results that the NNSD's, for all $k$ values, are very close to Wigner.  The variance of NNSD is found $\sigma^2(0) \sim 0.28$, for all $k$ values in EGOE($k$) and BEGOE($k$), which indicates the systems are completely chaotic with the unfolding procedure utilized here. Furthermore, the $\Delta_3$ statistic is obtained for unfolded spectrum using overlap interval of 2 and following the procedure in \cite{BHP} $\Delta_3(L)$ for $L \leq 60$ are calculated, where $L$ is the energy interval, measured in units of average level spacing, over which $\Delta_3(L)$ is calculated. Here also ensemble averaged results are compared with the Poisson and GOE forms. Once again it is seen that, with the given unfolded spectra, the $\Delta_3(L)$ statistic approaches the GOE values for all $k$ in EGOE($k$) and BEGOE($k$) examples.

\section{Conclusions}
\label{sec:6}

In this work, using the normal mode decomposition and the periodogram analysis, it is conclusively demonstrated that for many-particle quantum systems with $k$-body interactions, generically called the $k$-body the embedded Gaussian orthogonal random matrix ensemble EGOE($k$) for fermions (and BEGOE($k$) for bosons), exhibit a well defined average-fluctuation separation in the state density. The present analysis clearly demonstrates that the smooth part of state density is very well represented by the $q$-normal distribution ($f_{qN}$) with corrections defined by only few modes with long wavelengths and the remaining fluctuations are of GOE for all body rank $k$, both for fermion as well as boson systems. The effective number of modes contributing the smooth part decreases due to rapid decrease in intensity as the body rank $k$ increases and thus, GOE fluctuations set in faster.

Going beyond this, some studies have been reported for $k$-body interacting many-particle quantum systems in a mean field defined by non degenerate sp states generated by a one-body Hamiltonian ($h(1)$) by defining $H=h(1)+\lambda V(k)$, where $V(k)$ is EGOE($k$) or BEGOE($k$) \cite{Vyas2019,rao21}. For BEGOE(1+$k$), it is demonstrated that the smooth forms for the state density can be represented by $f_{qN}(E|q)$ for all $k$ values \cite{rao21}. For these ensembles, transition from order to chaos in NNSD is expected as $k$-body interaction strength $\lambda$ increases. Using the unfolding procedure described in the present work, the critical interaction strength $\lambda_c$, which defines the order to chaos transition, can be determined for given $k$-body interaction. With this, it is possible to establish that the embedded ensembles with $k$ body interactions are  generic models for strongly interacting quantum systems (fermions or bosons) in the chaotic domain. This will be addressed in future.
	
\section*{Acknowledgements}
Author thanks V. K. B. Kota and V. Potbhare for many
useful discussions. This work is a part of University
supported research project [grant No: GCU/RCC/2021-
22/20-32/508].

\end{document}